\documentclass[usenatbib]{mnras}
 
\usepackage[utf8]{inputenc}
\usepackage[T1]{fontenc}
\usepackage{ae,aecompl}

\usepackage{amsmath}
\usepackage{amssymb}
\usepackage{eqnarray}
\usepackage{graphicx}
\usepackage{tabularx}
\usepackage{placeins}
\usepackage{siunitx}
\usepackage{titlesec}
\usepackage{epstopdf}
\usepackage{flafter}
\usepackage{multirow}
\usepackage{appendix}

\usepackage{color}

\graphicspath{ {./Figures/} }

\title[A PCA of PAH emission]{A Principal Component Analysis of polycyclic aromatic hydrocarbon emission in NGC~7023}

\author[A. Sidhu et al.]{Ameek Sidhu$^{1,2}$\thanks{E-mail: asidhu92@uwo.ca}, Josh Bazely$^{1}$, Els Peeters$^{1,2,3}$, Jan Cami$^{1,2,3}$\\
$^{1}$Department of Physics \& Astronomy, University of Western Ontario, London, ON, N6A 3K7, Canada\\
$^{2}$Institute for Earth and Space Exploration, University of Western Ontario, London, ON, N6A 3K7, Canada\\
$^{3}$SETI Institute, 189 Bernardo Avenue, Suite 100, Mountain View, CA 94043, USA}

\begin{document}

\date{}

\maketitle

\begin{abstract}
We carried out a principal component analysis (PCA) of the fluxes of five polycyclic aromatic hydrocarbon (PAH) bands at 6.2, 7.7, 8.6, 11.0, and 11.2 $\mu$m in the reflection nebula NGC~7023 comprising of the photodissociation region (PDR) and a cavity. We find that only two principal components (PCs) are required to explain the majority of the observed variance in PAH fluxes (98\%). The first PC ($PC_{1}$), which is the primary driver of the variance, represents the total PAH emission.
The second PC ($PC_{2}$) is related to the ionization state of PAHs across the nebula. This is consistent with the results of a similar analysis of the PAH emission in NGC~2023. The biplots and the correlations of PCs with the various PAH ratios show that there are two subsets of ionic bands with the 6.2 and 7.7 $\mu$m bands forming one subset and the 8.6 and 11.0 $\mu$m bands the other. In addition, we further observe a distinction in the behaviour of the 8.6 and 11.0 $\mu$m bands. However, the distinction between these subsets is only present in the PDR. We have also carried out a separate PCA analysis of the PAH fluxes, this time only considering variations in the cavity. This shows that in the cavity, $PC_{2}$ is not related to the charge state of PAHs, but possibly to structural molecular changes. 
\end{abstract}


\section{Introduction}
\label{sec:Introduction}

Polycyclic Aromatic Hydrocarbons (PAHs) are a class of large organic molecules with carbon atoms arranged in a honeycomb-like structure and hydrogen atoms attached at the edges. These molecules are ubiquitous in the universe and are observed via their characteristic vibrational emission features in the mid-infrared (MIR) at 3.3, 6.2, 7.7, 8.6, 11.2, and 12.7 $\mu$m \citep[e.g.][]{Sellgren:83, Hony:oops:01, Peeters:prof6:02, Geers:06, SmithJD:07, Galliano:08}. In addition to the strong features, PAHs exhibit a plethora of weak features at 5.25, 5.65, 6.0, 6.7, 7.2-7.4, 8.2, 10.1, 10.5, 10.8, 11.0, 12.0,
13.5, 14.2, 15.8, 16.4, 17.4, and 17.8 $\mu$m \citep[e.g.][]{Allamandola:89, Moutou:2000, Peeters:2002, Peeters:2004, Werner:2004, Boersma:2009}. PAHs represent up to 15 \% of the cosmic carbon \citep{Tielens:2008}, however, the specific PAH molecules comprising the astronomical population is still largely unknown.

Subtle variations in relative intensities and profiles of PAH emission features observed in different environments have been found in the MIR studies of a large number of Galactic and extragalactic sources \citep[e.g.][]{Peeters:prof6:02, SmithJD:07, Sandstrom:2012, Matsuura:14, Shannon:2016, Peeters:17}. These variations are related to the physical conditions such as gas density, temperature, metallicity, and radiation field strength of environments where PAHs reside \citep{Galliano:08, Pilleri:2012, Boersma:15, Stock:17}. Comparison of astronomical observations with laboratory and theoretical studies have indicated that changes in physical conditions cause changes in PAH properties such as the ionization state, size, molecular structure, and molecular symmetry, resulting in the variations in PAH emission features \citep[e.g.][]{Hony:oops:01, Galliano:08, Bauschlicher:08, Bauschlicher:09, Ricca:12, Candian:14,  Boersma:16, Bouwman:2019}.

\citet{Sidhu:2021} analyzed the variability of five PAH emission features (at 6.2, 7.7, 8.6, 11.0, and 11.2 $\mu$m) in a Galactic Photodissociation region (PDR), NGC~2023 using a statistical technique called Principal Component Analysis (PCA). The feature at 11.0 $\mu$m is the weakest of these PAH features. It was chosen for PCA analysis because its ratio with the 11.2 $\mu$m band is a better tracer of the ionization state than the other ionized PAH bands \citep[e.g.][]{Rosenberg:11, Peeters:17}. This is due to the fact that the 11.0 $\mu$m band originates from the out of plane bending modes of solo C-H groups in ionized PAH molecules, whereas the 11.2 $\mu$m band originates from the same mode in neutral PAH molecules; thus, the ratio 11.0/11.2 traces only the ionization state of PAHs, with no dependence on any other molecular parameter \citep[e.g.][]{Hudgins:tracesionezedpahs:99, Hony:oops:01, Bauschlicher:08, Bauschlicher:09}. From the PCA analysis of PAH fluxes in NGC 2023, \citet{Sidhu:2021} concluded that the amount of PAH emission and the ionization fraction are the two key drivers of the observed PAH variations. Furthermore, the PCA analysis revealed a peculiar behaviour of the ionized PAH features at 6.2, 7.7, 8.6, and 11.0 $\mu$m. The features at 6.2 and 7.7 $\mu$m were found to form one group of ionic bands, while those at 8.6 and 11.0 $\mu$m form another distinct group. We argued that this distinction would arise if the 6.2 and 7.7 $\mu$m bands belong to less ionized PAHs, and the 8.6 and 11.0 $\mu$m to more ionized PAHs. In this paper, we extend the PCA analysis to another well known galactic PDR, NGC~7023, and compare and contrast the results with the previous study on NGC~2023. We will investigate whether the results obtained for NGC~2023 also hold for NGC~7023.

This paper is organized as follows. In Section~\ref{sec:NGC_7023}, we describe the source NGC~7023. In Section~\ref{sec:PCA} we provide a brief overview of PCA. We present the results of our PCA analysis of PAH emission in NGC~7023 in Section~\ref{sec:Results}, followed by a discussion on distinct PAH emission in the PDR and the cavity of NGC~7023 in Section~\ref{sec:cavity_PDR}. We discuss the origin of the subsets of ionic bands in Section~\ref{sec:ionic_bands} and compare the results of this study with a previous study on NGC~2023 in Section~\ref{sec:comparison}. Finally, we present a summary of our results in Section~\ref{sec:Summary}.

\section{NGC~7023}
\label{sec:NGC_7023}

\begin{figure}
\centering

\includegraphics[scale=0.35]{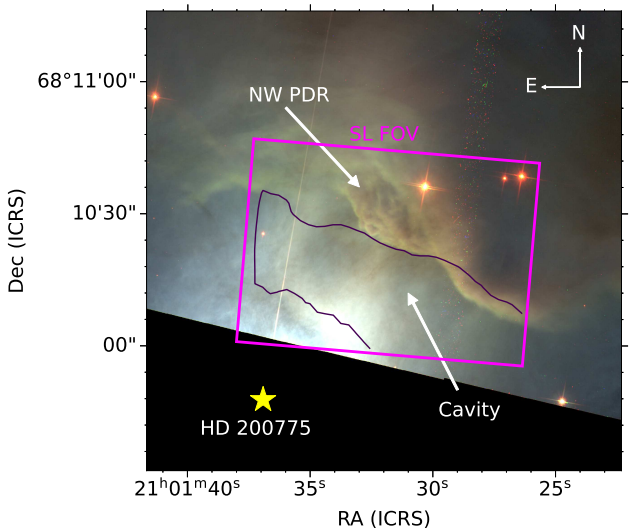}
\caption{The three-color Hubble Space Telescope ACS image of NGC~7023. The red channel corresponds to the combined optical H$\alpha$ (658 nm) and infrared I-band (850 nm) filters, the green channel to the optical V-band (625 nm) filter, and the blue channel to the optical B-band (475 nm) filter. The IRS-SL field of view (FOV) is shown in a pink rectangle. The yellow star indicates the position of the illuminating star HD~200775. The NW PDR and the cavity carved out by the star are also annotated in the figure. The contour of $PC_{2}$ = -0.24 separates the cavity and the PDR (see text in section~\ref{subsecec:spatial_maps} for details).}
\label{fig:IRAC_7023}
\end{figure}

NGC~7023 is a bright visual reflection nebula 430 pc away from the Earth, and illuminated by HD~200775, a spectroscopic binary (Herbig B3Ve - B5)  \citep{VandenAncker:1997, Witt:2006, Alecian:2013}. It is a well-studied reflection nebula owing to its high surface brightness and proximity to the Earth \citep[e.g.][]{Watt:1986, Chokshi:88, Sellgren:1992, Fuente:1993, Rogers:95, Lemaire:1996, Gerin:1998, An:2003, Witt:2006, Joblin:10, Rosenberg:11, Berne:12, Montillaud:2013, Kohler:2014, Boersma:15, Croiset:16, Le:2017, Joblin:18}. Observations in the ultraviolet, optical, and at infrared wavelengths show that the central star has carved out a cavity (gas density $\sim$ 100 cm$^{-3}$) in the nebula \citep[][see Fig.~\ref{fig:IRAC_7023}]{Watt:1986, Gerin:1998, Joblin:10, Berne:2015}. The walls of the cavity are surrounded by PDRs at $\sim$ 42$''$ North West (NW), 55$''$ South West (SW), and 150$''$ East (E) of the central star. These PDRs are made up of diffuse gas, $n_{H}$ $\sim$ $10^{3}$ - $10^{4}$ cm$^{-3}$, embedded with relatively dense filamentary structures ($n_{H}$ $\sim$ $10^{5}$ - $10^{6}$ cm$^{-3}$) of size $\sim$ 0.004 pc or less \citep{Chokshi:88, Rogers:95, Fuente:1996, Fuente:1999, Martini:1999, Kohler:2014, Joblin:18}. In this paper, we analyzed the emission from PAHs using the MIR data obtained with the Infrared Spectrograph \citep[IRS][]{Houck:04} in the Short-Low (SL) module (Spectral resolution $\sim$ 60-128, pixel scale $\sim$ 1.8$''$), on board the Spitzer Space Telescope \citep{Werner:04a}. Fig.~\ref{fig:IRAC_7023} shows the SL field of view (FOV) in which the NW PDR and a part of the cavity are observed.

\section{Principal Component Analysis}
\label{sec:PCA}
A Principal Component Analysis (PCA) is an unsupervised learning technique widely used in astronomy to analyse complex data sets \citep[e.g.][]{Wang:2011, Hurley:2012, Ensor:2017, Sidhu:2021}. PCA reduces the dimensionality of a data set by transforming the original set of variables into a new set of variables called the principal components (PCs) in such a way that only the first few PCs contain most of the statistical information about the data set \citep{Pearson:1901, Hotelling:1933, Jolliffe:16}. Here, we present a summary of PCA. A comprehensive mathematical formulation of PCA, is given in \citet{Sidhu:2021}.

To perform PCA, we first standardize the original data set so that each variable has a mean of zero and a standard deviation of one. PCs are then derived from the eigenvector decomposition of the covariance matrix of the standardized data set. These eigenvectors then form the PCs, with their relative importance determined by their corresponding eigenvalues. The eigenvector with the largest eigenvalue becomes the first PC and accounts for most of the variance in the data set. Each succeeding PC accounts for less variance than the preceding one. In this way, by transforming from original variables into PCs, we get, in principle, the parameters that drive the variance in the data set.

\section{PCA of PAH fluxes in NGC~7023}
\label{sec:Results}

\subsection{Measurement of PAH bands}
\label{subsecec:Measurement}

\begin{table}
    \centering
    
    \begin{tabular}{c c c}
    \hline
    \multirow{2}{*}{PAH band} & $\langle I_{PAH}\rangle$ & $\sigma_{PAH}$  \\
    & ($\times 10^{-5}$) & ($\times 10^{-6}$)\\
    \hline
    6.2 & 1.225 & 5.056  \\ 
    7.7 & 2.037 & 8.267 \\
    8.6 & 0.375  & 1.838 \\
    11.0 & 0.009  &  0.048\\
    11.2 & 0.460  & 2.899 \\
    \hline
    \end{tabular}
    \caption{The mean ($\langle I_{PAH}\rangle$) and standard deviation ($\sigma_{PAH}$) values of the PAH band flux variables in the SL FOV of NGC~7023. All values are in units of ${\rm W m}^{-2}{\rm sr}^{-1}$.}
    \label{tab:stats_flux}
\end{table}

We performed a PCA of the extinction corrected fluxes of the 6.2, 7.7, 8.6, 11.0, and 11.2 $\mu$m PAH bands observed in the SL FOV of NGC~7023. We obtained the flux measurements of these PAH bands from \citet{Stock:2016}. Here we briefly summarize their flux measurement strategy. First, they subtracted the continuum from the spectra by computing a spline fit to a set of continuum points at wavelengths of 5.37, 5.46, 5.86, 6.58, 6.92, 8.28, 9.15, 9.40, 9.64, 10.14, 10.33, 10.76, 11.82, 13.18, 13.49, 14.43, and 14.74 $\mu$m. Fluxes of the strong bands at 6.2, 7.7, 8.6, and 11.2 $\mu$m were then measured by direct integration. Fluxes of the 6.2 and 11.2 $\mu$m bands measured this way contained contributions from the weaker 6.0 and 11.0 $\mu$m bands, which they accounted for by fitting the Gaussian profiles at 6.0 and 11.0 $\mu$m to determine their contribution. The fluxes of 6.0 and 11.0 $\mu$m bands thus obtained were then subtracted from the fluxes of 6.2 and 11.2 $\mu$m bands measured with direct integration to obtain the final flux of the 6.2 and 11.2 $\mu$m bands. They calculated the uncertainties by comparing the integrated feature flux and the rms noise of featureless areas of the continuum between 9.3 and 9.5 $\mu$m, 13.3 and 13.5 $\mu$m, and 13.7 and 13.9 $\mu$m, respectively.

For our PCA analysis, we first masked the pixels where the signal-to-noise ratio (SNR) of fluxes in any of the five bands is less than 3. We note that the 11.0 $\mu$m band is the weakest of the five PAH bands considered in this work, and as a result, the SNR of the 11.0 $\mu$m band determines which pixels are masked. We also masked the pixels contaminated with diffraction effects from the central star and the two 2MASS point sources in our FOV (see Fig.~\ref{fig:IRAC_7023}). We then standardized the fluxes from the remaining pixels before using them as input variables in our PCA analysis. After the standardization operation, all the input flux variables have a mean of zero and a standard deviation of unity. The standardization is done so that all the input variables in PCA have comparable magnitudes that would result in meaningful PCs as output. We provide the mean and the standard deviation values of the PAH band flux variables used in our PCA analysis in Table \ref{tab:stats_flux}.

\subsection{Principal Components}
\label{subsec:Principal_components}

The five PCs, i.e. the unit eigenvectors, that result from our PCA are then given by these equations: 
\begin{equation}
\begin{split} 
PC_{1} = & \,\,  0.481\, z_{6.2} + 0.476\, z_{7.7} + 0.474\, z_{8.6}  \\
& + 0.461\, z_{11.0} + 0.322\, z_{11.2}\\
PC_{2} = & \,\,0.117\, z_{6.2} - 0.149\, z_{7.7} - 0.267 \,z_{8.6}  \\
& - 0.316\, z_{11.0} + 0.890\, z_{11.2}\\
PC_{3} =&\,\, - 0.396\, z_{6.2} - 0.575\, z_{7.7} + 0.184\,z_{8.6}  \\
& + 0.648\,z_{11.0} + 0.241\,z_{11.2}\\
PC_{4} =&\,\, - 0.699\, z_{6.2} + 0.647\, z_{7.7} - 0.198\,z_{8.6}  \\
& + 0.133\,z_{11.0} + 0.188\, z_{11.2}\\
PC_{5} =&\,\, - 0.331\, z_{6.2} - 0.039\, z_{7.7} + 0.794\,z_{8.6}  \\
& - 0.499\,z_{11.0} + 0.097\,z_{11.2}
\end{split}
\label{eq:PC_south}
\end{equation}

\noindent
where \{$z_{6.2}$, $z_{7.7}$, $z_{8.6}$, $z_{11.0}$, $z_{11.2}$\} are the standardized flux variables.

Following \citet{Ensor:2017}, we used a a Monte Carlo Simulation to estimate the uncertainties of the coefficients of the standardized flux variables in PCs (equation \ref{eq:PC_south}). We generated 1000 new data sets by adding random noise to the original PAH flux variables. For each data set, we selected a random noise from the normal distribution of PAH flux variables with a mean of zero and standard deviation equal to the error on the flux measurement. We then standardized the new data sets and performed a PCA on each of the new standardized data sets resulting in 1000 eigen vectors for each PC. We then considered the highest computed value of the coefficients of the standardized flux variables in PCs as the upper limit on the coefficients and the lowest computed value as the lower limit on the coefficients of the standardized flux variables in PCs. We provide the upper and lower limits on the coefficients in Appendix~\ref{sec:uncertainties}.

\begin{table}
    \centering
    \begin{tabular}{c c c c}
    \hline
    \multirow{1}{*}{PC} & \% variance explained & Uncertainty interval \\
    \hline
    1 & 83.9 & 82.6 -- 83.9 \\ 
    2 & 14.1 & 13.8 -- 14.5 \\
    3 & 1.7 & \,\,\,1.7 -- \,\,\,2.5\\
    4 & 0.2 & \,\,\,0.2 -- \,\,\,0.9 \\
    5 & 0.1 & \,\,\,0.1 -- \,\,\,0.2\\
    \hline
    \end{tabular}
    \caption{Fraction of variance explained by the principal components (PCs). The last column represents the uncertainty in the \% variance explained by the PCs estimated by the Monte Carlo Simulation.}
    \label{tab:percent_variance}
\end{table}

The relative importance of these PCs is determined by the amount of variance explained by each PC. Table \ref{tab:percent_variance} lists the fraction of the variance explained by the PCs obtained from our PCA analysis. Recall that the fraction of variance explained by a PC is the eigenvalue of that PC eigenvector. We note that the first two PCs combined explain $\sim$98\% of the variance in the data, with the first PC explaining the majority of the variance. Therefore, the first two PCs are the primary drivers of the PAH flux variations observed in NGC~7023. If we ignore the three remaining marginal PCs, we can then decompose the standardized flux variables into $PC_{1}$ and $PC_{2}$ as follows:

\begin{equation}
\begin{split} 
z_{6.2} = & \,\,  0.481\, PC_{1} + 0.117\, PC_{2} \\
z_{7.7} = & \,\,0.476\, PC_{1} - 0.149\, PC_{2} \\
z_{8.6} =&\,\, 0.474\, PC_{1} - 0.267\, PC_{2} \\
z_{11.0} =&\,\, 0.461\, PC_{1} - 0.316\, PC_{2} \\
z_{11.2} =&\,\, 0.322\, PC_{1} + 0.890\, PC_{2} 
\end{split}
\label{eq:std_var_decomposition}
\end{equation}

In the following sections, we investigate whether we can assign a physical interpretation to $PC_{1}$ and $PC_{2}$. For the remainder of our paper, we exclude the last three PCs explaining $\sim$2\% of the variance from any further analysis.

\subsection{PCA biplots}
\label{biplots}
\begin{figure}
\centering

\includegraphics[scale=0.42]{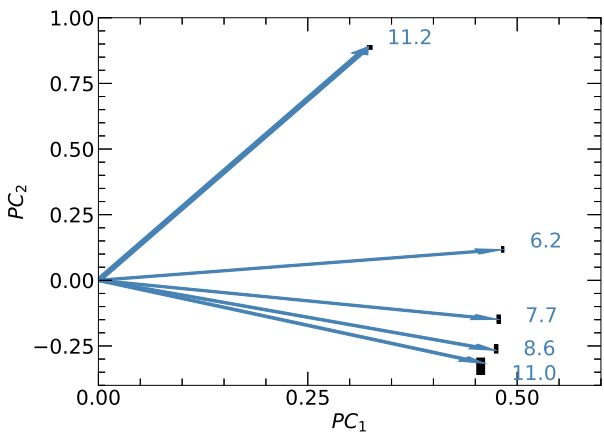}
\caption{Biplots showing projection of the standardized flux variables in $PC_{1}$-$PC_{2}$ plane in NGC~7023. The black rectangles indicate the uncertainty in the projection of PAH flux variables on PCs obtained via Monte Carlo Simulation.}
\label{fig:biplots}
\end{figure}

To interpret the physical meaning of $PC_{1}$ and $PC_{2}$, we begin by constructing biplots that depict the contribution of standardized flux variables to the PCs. Fig.~\ref{fig:biplots} shows the biplots obtained from our PCA analysis of PAH emission in NGC~7023 where we show the projection of the standardized flux variables in the reference frame of PCs. These projections are a measure of the correlation between the standardized flux variables and PCs, i.e. the larger the projection of the standardized flux variable on a PC, the larger the correlation of that variable with the PC. All the PAH bands considered here have a positive projection on $PC_{1}$: the traditional ionized PAH bands (at 6.2, 7.7, 8.6, and 11.0 $\mu$m) have a projection of $\sim$ 0.5 and the neutral PAH band (at 11.2 $\mu$m) has a slightly smaller projection of $\sim$ 0.3. On the other hand, the projections of PAH bands on $PC_{2}$ are more diverse. While the 7.7, 8.6, and 11.0 $\mu$m bands have a small negative projection ($\sim$ -0.1 -- -0.3) on $PC_{2}$, the 6.2 $\mu$m band exhibits a small positive projection ($\sim$ 0.1), and the 11.2 $\mu$m band a large positive projection ($\sim$ 0.9) on $PC_{2}$.

The trends observed in the biplots suggest that $PC_{1}$ represents the characteristic PAH emission of a mixture of neutral and ionized PAHs. The slight difference in the projections of the ionized and neutral PAH bands on $PC_{1}$ suggests that the characteristic PAH emission in NGC~7023 has more contribution from ionized PAHs than neutral PAHs. Directions of the PAH band projections on $PC_{2}$ distinguishes between the neutral and the ionized PAH bands with the neutral band (at 11.2 $\mu$m) exhibiting a positive projection and the ionized bands (at 7.7, 8.6, and 11.0 $\mu$m except at 6.2 $\mu$m) exhibiting a negative projection, thereby indicating that $PC_{2}$ is related to the ionization state of PAHs. The different behaviour of the 6.2 $\mu$m PAH band compared to the other ionized PAH bands at 7.7, 8.6, and 11.0 $\mu$m is worth mentioning as it depicts the subtle differences between the ionic bands which have traditionally been treated similarly. We address these differences in ionic bands further in Section \ref{sec:ionic_bands}.

\subsection{Spatial Maps of PCs}
\label{subsecec:spatial_maps}
\begin{figure*}
\centering
\begin{tabular}{cc}
    \includegraphics[angle=265.11918, scale=0.45]{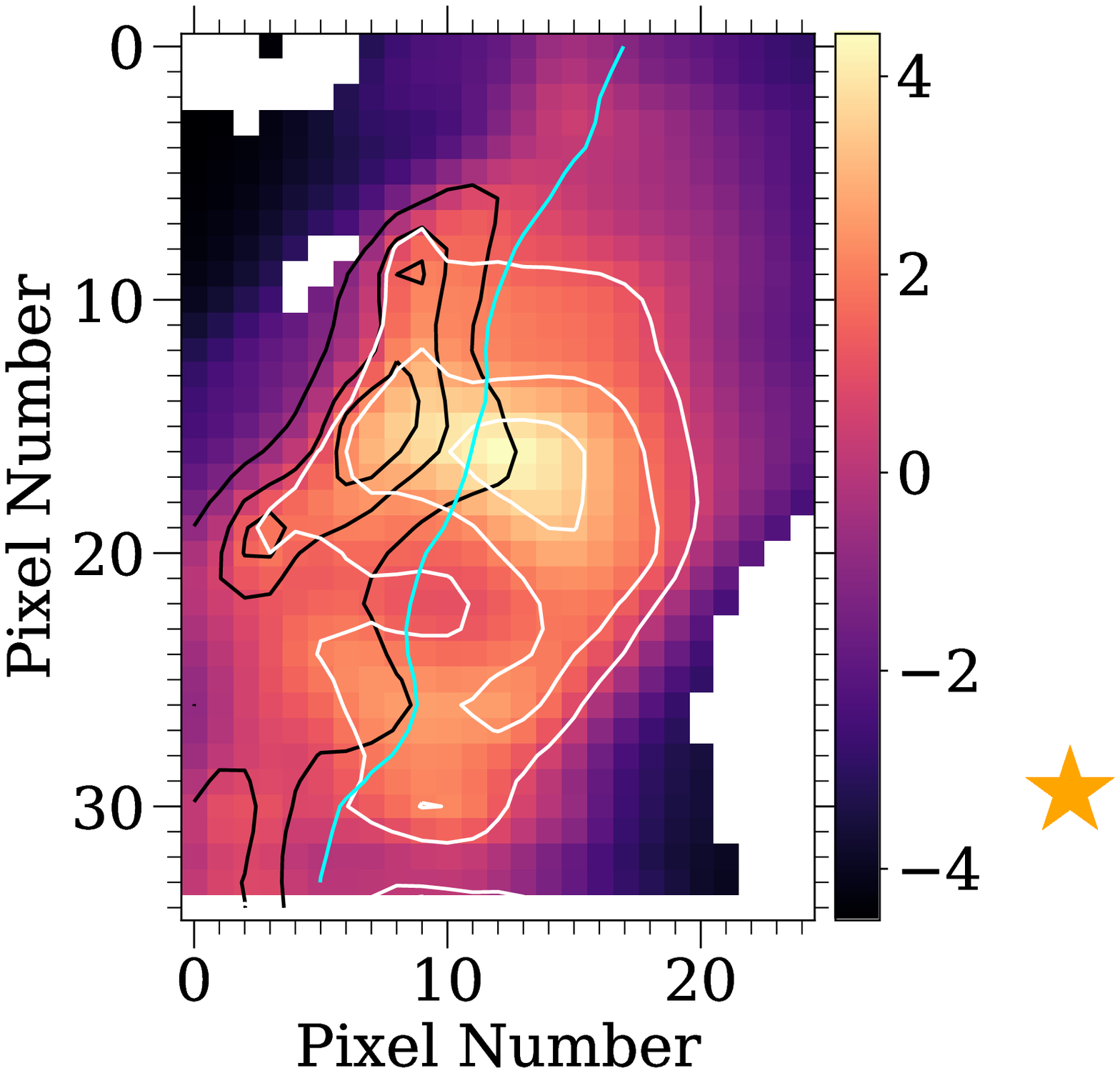} & \includegraphics[angle=265.11918, scale=0.45]{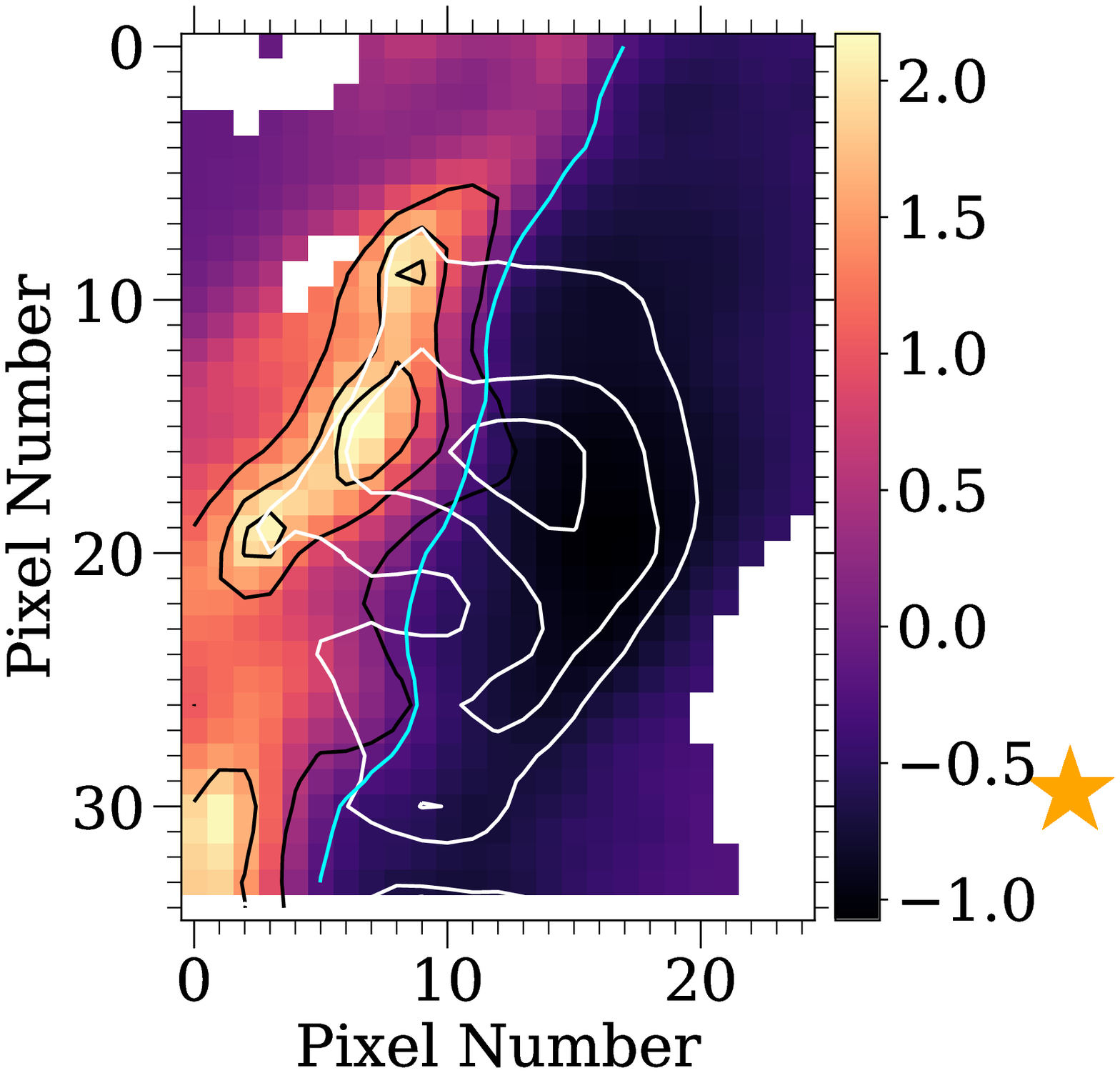}  \\
     & 
\end{tabular}
 
\caption{Spatial maps of $PC_{1}$ (left) and $PC_{2}$ (right) in the SL FOV of NGC~7023. For reference, the contours of the 7.7, 11.2 $\mu$m PAH intensity, and $PC_{2}$ = -0.24 are overlaid in white, black, and cyan respectively. Pixels where the SNR of the fluxes in any of the five PAH bands used in PCA is less than 3, and those contaminated with the diffraction effects from the illuminating star and the 2MASS point sources in the FOV are shown in white. The yellow star outside the FOV indicates the position of the illuminating star.}
\label{fig:spatial_map_PCs}

\end{figure*}

We also studied the spatial distribution of the magnitude of the PC eigenvectors in order to better understand their physical meaning. We emphasize that although the PC eigenvectors given by equations \ref{eq:PC_south} have unit magnitude, their magnitude is not unity across the spatial map of the nebula. In Fig.~\ref{fig:spatial_map_PCs} we present the spatial maps of $PC_{1}$ and $PC_{2}$ overlaid with the contours of the 7.7 and 11.2 $\mu$m PAH intensities. The star is at the bottom right corner of those maps. High $PC_{1}$ values (yellowish hues) form a ring-shaped structure which is also traced well by the 7.7 $\mu$m PAH band (white contours). The ring-shaped structure was first identified by \citet{An:2003} in their intensity map of the 3.29 $\mu$m PAH band. The origin of this ring-shaped structure is still unknown. Intermediate $PC_{1}$ values (purple hues) form an elongated rectangular patch and belong to regions outside the ring-shaped structure. Low $PC_{1}$ values (dark) are further present in regions outlining the rectangular patch of intermediate $PC_{1}$ values, in the lower right and top left corner of the map. We note that the spatial morphology of $PC_{1}$ looks identical to the spatial distribution of total PAH flux (see figure \ref{fig:spatial_map_total_PAH_flux}) thereby reinforcing our earlier suggestion that $PC_{1}$ represents the characteristic PAH emission in NGC~7023.

The spatial map of $PC_{2}$ is quite intriguing as it offers insight into the PAH emission characteristics of NGC~7023 beyond the general picture that emerges from the biplots in Section~\ref{biplots}. We note that the high $PC_{2}$ values ($\gtrsim$1.5) are co-located with the strong 11.2 $\mu$m PAH emission (shown in the black contour map). The 11.2 $\mu$m PAH emission aligns very well with the H$_2$ emission tracing the PDR front \citep{Berne:2009, Croiset:16}. On the other hand, low $PC_{2}$ values ($\lesssim$-0.24) originate from the cavity and intermediate $PC_{2}$ values ($\sim$ -0.24 -- 1.5) probe the transition from the PDR to the cavity as well as the region behind the PDR ridge, i.e. the top left corner of the map. The ring-shaped structure is not visible in the spatial map of $PC_{2}$ though the lowest $PC_{2}$ values (black color) are found slightly offset (to the right, i.e. in the direction of the central star) from the peak of the PAH emission found in the ring-shaped structure and thus not closest to the central star. Thus, the spatial distribution of $PC_{2}$ clearly distinguishes between the cavity and the PDR.  In the framework of PCA, $PC_{2}$ is a first-order correction to the characteristic PAH emission traced by $PC_{1}$. Therefore, the differentiation between the cavity and the PDR highlighted by $PC_{2}$ implies that the PAH properties vary between these two regions. Furthermore, since $PC_{2}$ is related to the ionization state of PAHs (see Section~\ref{biplots}), PAHs in the cavity and the PDR seems to differ primarily in their ionization states. 

We compared our results to those of \citet{Rapacioli:2005}, who used Singular Value Decomposition to analyze the mid-IR ISOCAM spectral maps of NGC 7023. These authors extracted four distinct signals corresponding to PAH cations, PAH neutrals, carbonaceous very small grains, and the continuum. The spatial distribution of these signals revealed that neutral PAHs were more prominent in the PDR, whereas PAH cations were more prominent near the star. The PCA analysis performed in this paper also reveals that the PAHs in the PDR and the cavity differ primarily in terms of their ionization state. The comparison of the spatial maps of PCs with the results of \citet{Rapacioli:2005} therefore indirectly strengthens our interpretation of PCs.


 



 





\subsection{Correlation plots}
\label{subsec:correlation_plots}
\begin{figure*}
\centering

\resizebox{\hsize}{!}{\includegraphics{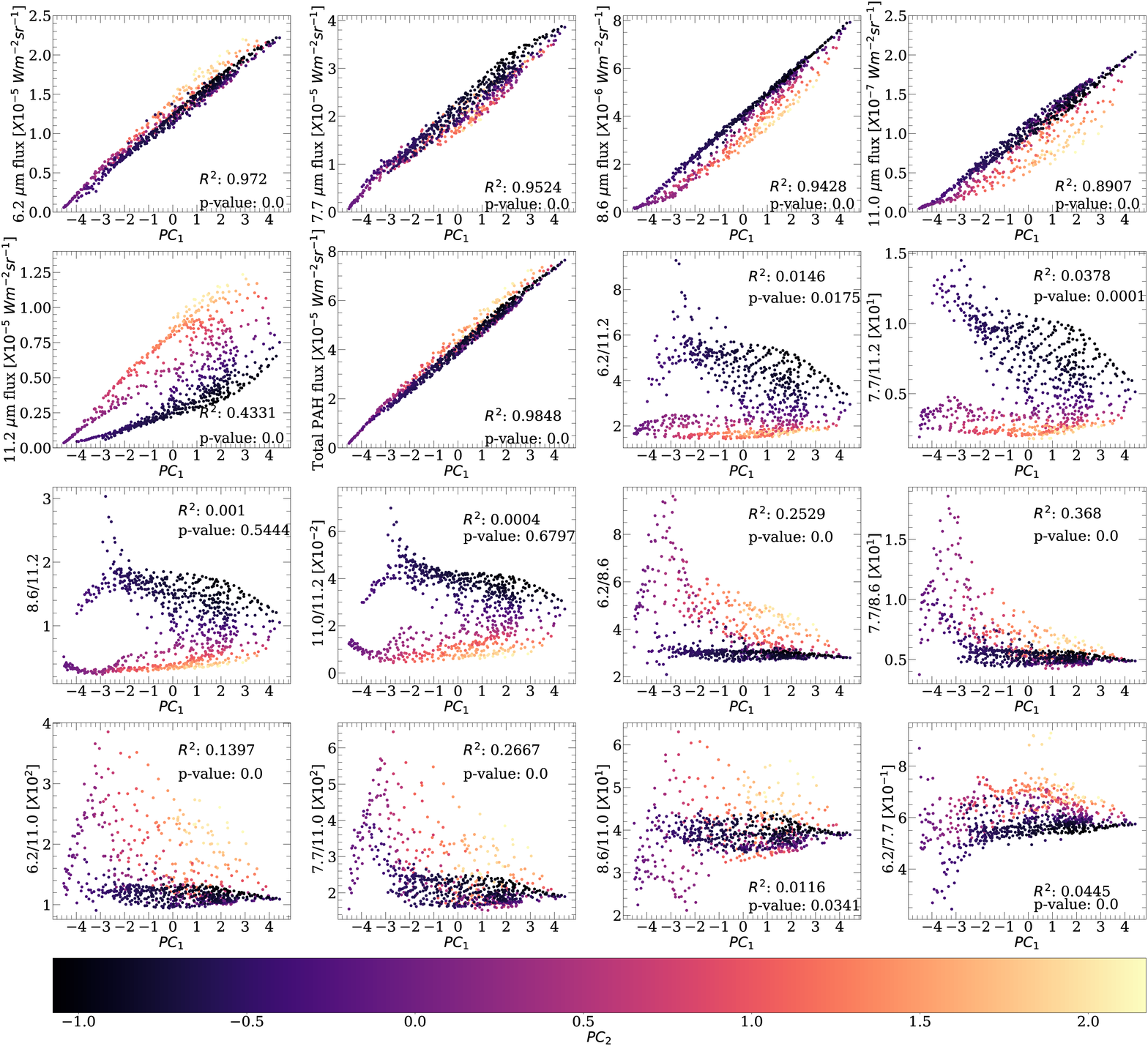}}

\caption{Correlations of $PC_{1}$ with the PAH fluxes and the PAH ratios, colour-coded with $PC_{2}$ values. The Pearson correlation coefficient and the corresponding p-value of statistically independent data points are shown in the corner of the each plot (see text for details)}.
\label{fig:correlations_PCs}

\end{figure*}
We further analysed the correlations of PCs with the PAH fluxes and the PAH ratios. Since $PC_{1}$ and $PC_{2}$ jointly account for the majority of the variance in PAH fluxes, we did not analyse the $PC_{1}$ and $PC_{2}$ correlations separately. Instead, we studied the correlations of both PCs together by plotting the correlations of $PC_{1}$ with the PAH fluxes and the PAH ratios and colour-coding the data points with their $PC_{2}$ values (see Fig.~\ref{fig:correlations_PCs}). For each of the correlation, we calculated the Pearson correlation coefficient ($R^{2}$) and the corresponding p-value of statistically independent data points only. For calculating the correlation coefficient and p-value, we only took the data from every other pixel, i.e. leaving a pixel to the left, right, top, and bottom of the selected pixel.

First of all, we note that $PC_{1}$ exhibits a linear relationship with the individual PAH fluxes and the total PAH flux (i.e. the sum of the fluxes of PAH bands at 6.2, 7.7, 8.6, 11.0, and 11.2 $\mu$m) but does not correlate with any of the PAH ratios. Moreover, the best correlation of $PC_{1}$ is with the total PAH flux with a correlation coefficient of 0.9851, thus lending support to our arguments in Sections~\ref{biplots} and \ref{subsecec:spatial_maps} that $PC_{1}$ probes the characteristic PAH emission in NGC~7023.

Secondly, there are observable branches in the $PC_{1}$-11.2 $\mu$m plot, which are neatly separated by their $PC_{2}$ values. Such branches are also evident in the $PC_{1}$-8.6 and 11.0 $\mu$m plots, although not as cleanly as observed in the $PC_{1}$-11.2 $\mu$m plot. In the branches of $PC_{1}$-11.2 $\mu$m plot, we note that for a given $PC_{1}$ value, low $PC_{2}$ values corresponds to low 11.2 $\mu$m fluxes which gradually increases with increasing $PC_{2}$ values. These branches indicate different relationships between $PC_{1}$ and the 11.2 $\mu$m flux in our FOV, implying that the PAH populations leading to these distinct branches are different. Since these branches correspond to different $PC_{2}$ values and the fact that $PC_{2}$ clearly distinguishes between the cavity and the PDR as observed in the spatial map of $PC_{2}$ (see Section~\ref{subsecec:spatial_maps}), we once again reach the same conclusion that the PAH population in the cavity differs from that in the PDR. Moreover, it is clear that this difference in population is primarily driven by a difference in ionization fraction. Indeed, the data points for which $PC_{2} > -0.24 $ belong to the PDR and the transition region, and are characterized by low 6.2/11.2, 7.7/11.2, 8.6/11.2, and 11.0/11.2 PAH ratio, in contrast to the much higher ratios for data points originating from the cavity ($PC_{2}$ values $< -0.24 $). 

Finally, the correlation plots also illustrate the different character of the ionic bands in the PDR. We observe that the data points originating from the PDR and the regions surrounding it exhibit a broad range of values in the 6.2/8.6, 7.7/8.6, 6.2/11.0, 7.7/11.0, and 8.6/11.0 PAH ratios. The 6.2/7.7 PAH ratio, however, shows only a narrow range of values (see Appendix~\ref{sec:distribution_PAH_ratios_PDR} for the quantitative justification). This means that while the relative strengths of the 6.2 and 7.7 $\mu$m bands do not change significantly, there is a considerable change in the strengths of the 6.2 and 7.7 $\mu$m bands with respect to the 8.6 and 11.0 $\mu$m bands. Thus, the 6.2 and 7.7 $\mu$m bands form one subset of ionic bands that change together, but that behave differently from the other subset formed by the 8.6 and 11.0 $\mu$m bands.

\section{Comparing PAH emission in the PDR and the cavity of NGC~7023}
\label{sec:cavity_PDR}

\begin{figure}
\centering
\includegraphics[scale=0.42]{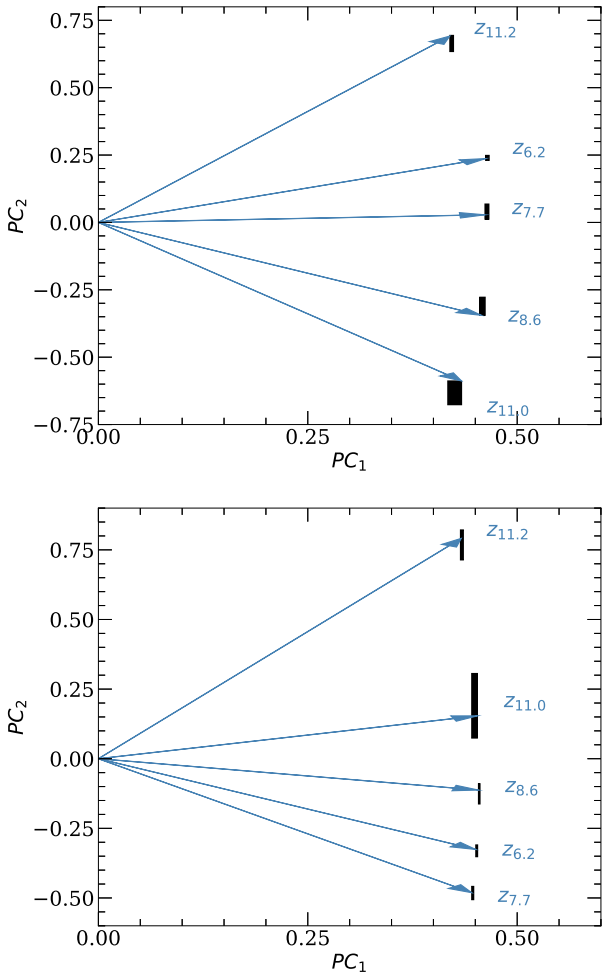}
\caption{Biplots of PCA in the PDR (top) and cavity (bottom) of NGC~7023. The black rectangles indicate the uncertainty in the projection of PAH flux variables on PCs obtained via Monte Carlo Simulation.}
\label{fig:biplots_PDR_cavity}
\end{figure}

The spatial map of $PC_{2}$ and the correlation analysis of PCs in Sections~\ref{subsecec:spatial_maps} and \ref{subsec:correlation_plots} respectively indicated that the PAH population in the cavity differs from the PDR. These results also highlighted the obvious difference between the ionization states of PAHs in the two regions. Furthermore, we discriminate two different subsets of the ionic bands that are evident only in the PDR environment and not in the cavity. In this section, we now further explore the variations in the PAH emission of the cavity and the PDR by performing two independent PCA analyses of the PAH emission in these two regions. We used the spatial map of $PC_{2}$ as the reference point to identify the data points originating from the PDR and the cavity, i.e. we assumed that the regions with $PC_{2} < -0.24$ belong to the cavity and the regions with $PC_{2} > -0.24$ belong to the PDR. We note that the regions with $PC_{2} > -0.24$ also includes the region separating the cavity from the PDR. Since the trends observed in the region separating the cavity and the PDR are similar to those observed in the PDR (Section \ref{subsec:correlation_plots}), we will treat this region identical to the PDR for the remainder of this paper.

 \begin{table}
    \centering
    \begin{tabular}{c c c}
    \hline
    \multirow{2}{*}{PC} & \multicolumn{2}{c}{\% variance explained} \\
    & PDR & Cavity\\
    \hline
    1 & 90.7 (88.8 -- 90.7) & 96.5 (95.4 -- 96.5) \\
    2 & 7.7 (7.2 -- 8.9) & 2.9 (2.8 -- 3.2)\\
    3 & 1.3 (1.3 -- 1.9) & 0.4 (0.4 -- 1.3) \\
    4 & 0.2 (0.2 -- 0.6) & 0.1 (0.1 -- 0.2) \\
    5 & 0.1 (0.1 -- 0.2) & 0.1 (0.1 -- 0.2) \\
    \hline
    \end{tabular}
    \caption{Fraction of variance explained by PCs obtained from two independent PCA of PAH emission in the PDR and the cavity of NGC~7023. The uncertainty interval in the \% variance explained estimated by the Monte Carlo simulation is given in brackets. The segregation of the PDR and the cavity is based on the spatial map of $PC_{2}$ obtained from the PCA analysis of PAH emission of the entire SL FOV of NGC~7023 (see text for details).}
    \label{tab:percent_variance_PCA_cavity_PDR}
\end{table}

The equations of the PCs resulting from the independent PCA analyses of the PDR and the cavity are given in Appendix \ref{sec:PCA_PAH_PDR} and \ref{sec:PCA_PAH_cavity} respectively. The amount of variance explained by each PC is presented in Table~\ref{tab:percent_variance_PCA_cavity_PDR} and shows that once more, the first two PCs ($PC_{1}$ and $PC_{2}$) account for the majority of the variance ($\sim$ 98-99 \%) in the PAH emission of both the PDR and the cavity. However, the fraction of the variance explained by individual PCs varies. For instance, $PC_{1}$ in the PDR accounts for $\sim$ 91 \% of the variance in the PAH emission compared to $\sim$ 96 \% in the cavity. $PC_{2}$, on the other hand, accounts for $\sim$ 7 \% of the variance in the PDR and only $\sim$ 3 \% in the cavity. Nonetheless, similar to the results of PCA on the entire SL FOV (see Section~\ref{subsec:Principal_components}), only two PCs are relevant to explain the variance in PAH emission in both the PDR and the cavity.

\subsection{Biplots in the PDR and the cavity of NGC~7023}
We then investigated the characteristics of the new PCs found in the PDR and the cavity by analysing the biplots for the PCs obtained in the two regions (see Fig.~\ref{fig:biplots_PDR_cavity}). For both the PDR and the cavity, all PAH bands have similar positive projections on $PC_{1}$, implying that $PC_{1}$ represents emission from a mixture of neutral and ionized PAHs. The projections of $PC_{2}$, on the other hand, are quite distinct between the PDR and the cavity. In the PDR, the projections of 11.0 and 11.2 $\mu$m PAH bands form the two ends of the spectrum exhibiting the largest negative and positive projection, respectively. The 6.2 and 7.7 $\mu$m bands have positive projections, with the projection of 6.2 $\mu$m band being greater than that of 7.7 $\mu$m band. The 8.6 $\mu$m PAH band, on the other hand, has a large negative projection although less than that of 11.0 $\mu$m PAH band. Since $PC_{2}$ provides a clear distinction between the ionized 11.0 $\mu$m band and the 11.2 $\mu$m neutral band, we conclude that $PC_{2}$ is describing the effect of changes in the charge state in the PDR. Furthermore, we note that the projections of the 6.2 and 7.7 $\mu$m ionized PAH bands follow the 11.2 $\mu$m band while the 8.6 $\mu$m ionized band follow the 11.0 $\mu$m band, implying that the 6.2 and 7.7 $\mu$m bands form a subset of ionic bands which behave differently than the 8.6 and 11.0 $\mu$m bands. 

In the cavity on the other hand, the projections of the 11.2 and the 7.7 $\mu$m PAH bands onto $PC_2$ are on opposite ends. However, the projection of the 11.2 $\mu$m band is twice as large as the 7.7 $\mu$m band. The 11.0 $\mu$m band exhibits a small projection in the direction of the 11.2 $\mu$m band while the 6.2 and 8.6 $\mu$m bands exhibit negative projections though less so than the 7.7 $\mu$m band. Thus, it appears that in the cavity, $PC_{2}$ is tracing some PAH property other than the ionization state. Interestingly, the positive projections correspond to PAH bands that represent solo C-H out of plane bending modes while the negative projections correspond to C-C stretching modes and combinations of C-H in plane bending modes \citep[from all C-H groups: solo's, duo's, trio's, ...; see e.g.][]{Hony:oops:01, Bauschlicher:09}. $PC_{2}$ could thus be probing changes in the molecular structure or hydrogenation states of PAHs. 
Hence, the projections of PAH bands on $PC_{2}$ in the PDR and the cavity suggest an underlying difference in photochemical evolution of PAHs in these two regions.

\subsection{Characteristic PAH emission spectrum in the PDR and the cavity of NGC~7023}
\begin{figure}
\centering

\includegraphics[scale=0.54]{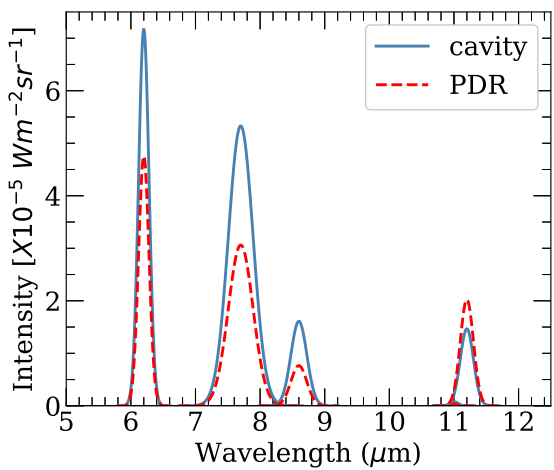}
\caption{Characteristic PAH spectrum of $PC_{1}$ in the PDR and cavity}
\label{fig:characteristic_spec_PC1}
\end{figure}



In Fig.~\ref{fig:characteristic_spec_PC1} we present the characteristic PAH spectrum of $PC_{1}$ representing the PAH emission in the PDR and the cavity. We followed the procedure given in \citet{Sidhu:2021} to derive the characteristic PAH spectrum of $PC_{1}$. First, we obtained the standardized fluxes for the PAH bands by substituting $PC_{1}$ = 1 and $PC_{2}$ = 0 in Equations \ref{eq:std_var_decomposition_PDR} and \ref{eq:std_var_decomposition_cavity}. To the standardized fluxes thus obtained we then applied the inverse standardization operation, i.e. we added the mean value of the original PAH flux variables in the cavity and the PDR to the product of the standard deviation of the original PAH flux variables and $z_{PAH}$. We find that the 6.2 and 7.7 $\mu$m emerge as strong bands in both the cavity and the PDR and that the 6.2, 7.7 and 8.6 $\mu$m bands are stronger in the cavity than in the PDR. The 11.2 $\mu$m band on the other hand is stronger in the PDR than in the cavity, implying that as we go from the cavity to the PDR, the contribution of neutral PAH molecules to the total PAH emission increases. The 11.0 $\mu$m band is a weak feature in both the cavity and the PDR.


\subsection{Comparing the PDR to the cavity: summary}
The independent PCA analysis of PAH emission in the PDR and the cavity thus establishes the following facts:
\begin{enumerate}
    \item The PAH population in the PDR is different from that in the cavity. The primary difference lies in the ionization fraction: the PAH population in the PDR contains more neutral PAH molecules while that in the cavity contains more ionized PAH molecules. 
    \item The photochemical evolution of PAHs is different between the PDR and the cavity. Variations in the PDR can be primarily ascribed to changes in the charge state; in the cavity, another molecular property drives much of the variations. 
    \item The peculiar behaviour of the ionic bands, i.e. the distinct behaviour of the 6.2 and 7.7 $\mu$m bands from the 8.6 and 11.0 $\mu$m bands, is only seen in the PDR and not in the cavity.
\end{enumerate}

\section{Subsets of the ionic bands and their behaviour}
\label{sec:ionic_bands}
\begin{figure}
\centering
\resizebox{\hsize}{!}{\includegraphics[scale=0.42]{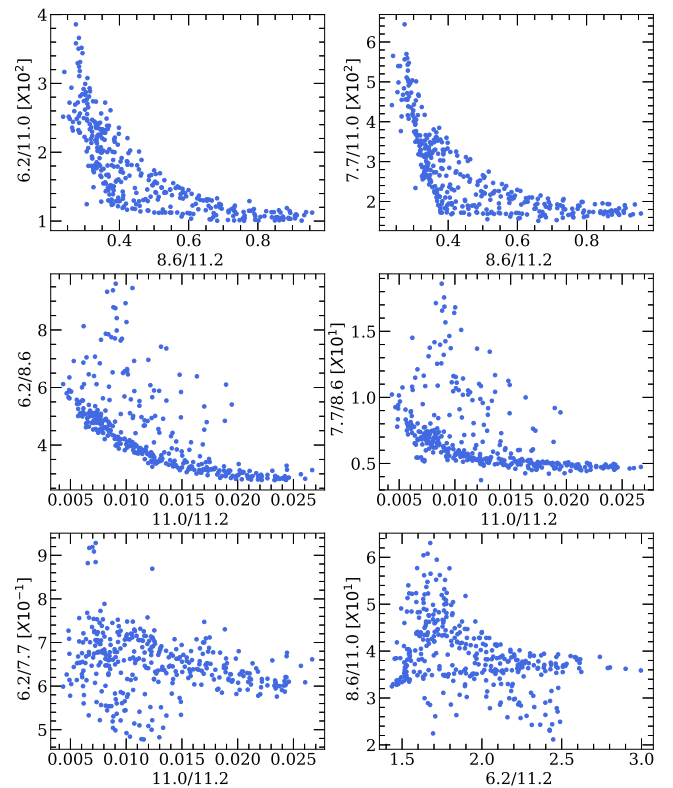}}

\caption{Correlations of ionic PAH ratios with the PAH ratios tracing ionization state} in the PDR of NGC~7023.
\label{fig:corr_subset_ionic_bands}
\end{figure}

The results of the PCA using our entire data set showed that the ionic PAH bands form two distinct subsets. The 6.2 and 7.7 $\mu$m bands form one subset that behaves differently than the other subset comprised of the 8.6 and 11.0 $\mu$m bands. 
This distinction is furthermore only present in the PDR and not in the cavity. Here, we will discuss the possible origin of this behaviour. 

In the biplots stemming from the PCA applied to the PDR only (see the top panel of Fig.~\ref{fig:biplots_PDR_cavity}), the 6.2 and 7.7 $\mu$m bands follow the 11.2 $\mu$m whereas the 8.6 and 11.0 $\mu$m bands point the other way. This suggests that the carrier of the 6.2 and 7.7 $\mu$m bands may be less ionized than the carrier of the 8.6 and the 11.0 $\mu$m bands. To test this hypothesis, we studied how the ratios of the ionic PAH bands correlate with the PAH ratio tracing the charge state of PAHs, in the PDR only (Fig.~\ref{fig:corr_subset_ionic_bands}). In each panel of Fig.~\ref{fig:corr_subset_ionic_bands}, we plot a PAH ratio tracing charge state on the x-axis and a ratio of the ionic PAH bands on the y-axis. The presence of a relationship in the 6.2/11.0 and 7.7/11.0 vs 8.6/11.2 and in the 6.2/8.6 and 7.7/8.6 vs 11.0/11.2 plots suggests that 6.2 and 7.7 $\mu$m bands probe a different ionization state than the 8.6 and 11.0 $\mu$m bands. Furthermore, the lack of a relationship between the 6.2/7.7 vs 11.0/11.2 plots and the 8.6/11.0 vs 6.2/11.2 plots indicates that there is significantly less distinction between the 6.2 and 7.7 $\mu$m bands and the 8.6 and 11.0 $\mu$m bands based on ionization state. Within this framework, the 6.2 and 7.7 $\mu$m bands are attributed to less ionized PAHs than the 8.6 and 11.0 $\mu$m PAH bands given that high values of the PAH ratio tracing charge state correspond to lower values of the 6.2/11.0, 7.7/11.0, 6.2/8.6, and 7.7/8.6 ratios.

Alternatively, the two different subsets of the ionic PAH bands could be the result of contributions from larger species such as PAH clusters or very small grains (VSGs) to the PAH bands in the 6-9 $\mu$m region. In their analysis of PAH emission in the 7-9 $\mu$m region in NGC~2023, \citet{Peeters:17} concluded that there are two distinct subpopulations contributing to the emission in this region. The spatial morphology of one subpopulation bore similarity with the 11.0 $\mu$m band and the other with the dust continuum and 5--10 $\mu$m PAH plateau. Several studies attribute the emission from plateaus to the PAH clusters or VSGs \citep[e.g.][]{Allamandola:89, Bregman:89, Peeters:17}. In this scenario, the subpopulation bearing similarity with the plateau emission could therefore contain a contribution from larger species such as PAH clusters or VSGs which in turn could be at the origin of the distinct nature of the ionic bands. Since the larger species can not survive in the cavity, they can not contribute to the ionic bands in the 6-9 $\mu$m region resulting in the absence of distinction between the ionic bands in the cavity. In contrast, in the more shielded PDR environments, PAH clusters and VSGs can survive and contribute to the fluxes of PAH features in the 6-9 $\mu$m region. The amount of contribution from larger species to the fluxes of 6.2, 7.7, and 8.6 $\mu$m bands would lead to the different characters of the ionic bands. In this regards, the amount of contribution from larger species to the 6.2 and 7.7 $\mu$m bands would be comparable and more than to the 8.6 $\mu$m band.

Finally, we note that the PAH properties other than charge such as size, molecular structure, and symmetry can also influence the behaviour of the ionic bands. However, based on our current state of knowledge of PAH astrophysics, we cannot systematically analyse the effect of these properties on the distinct behaviour of the ionic bands.

\section{Comparing the PAH PCA of NGC~2023 and NGC~7023}
\label{sec:comparison}

\begin{figure}
\centering
\includegraphics[scale=0.54]{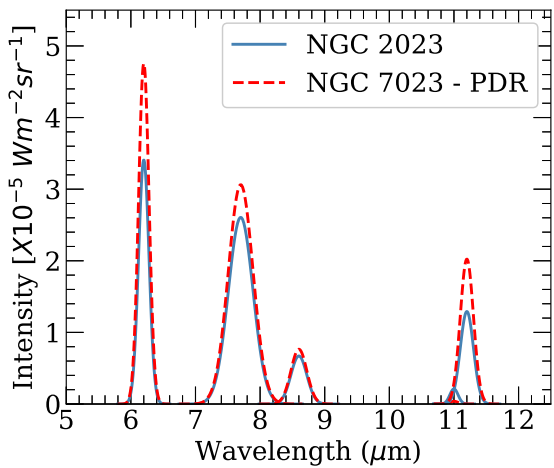}

\caption{Characteristic PAH spectrum of $PC_{1}$ in NGC~2023 and the PDR of NGC~7023.}
\label{fig:characteristic_spec_comparison}
\end{figure}

One of the goals of this paper is to compare and contrast the results of our PCA of PAH emission in NGC~7023 to a similar study on NGC~2023 \citep{Sidhu:2021} to investigate whether the results obtained pertaining to PAH emission are universal. The PCA analysis of PAH emission is performed in both environments using observations in the SL FOVs. While in NGC~7023, we studied one FOV comprising the dust-free cavity created by the exciting star (HD~200775) and the NW PDR; in NGC~2023, \citet{Sidhu:2021} investigated two FOVs north and the south of the exciting star (HD~37903), comprising of the PDRs embedded with bright ridges. These regions represent different physical conditions. In particular, while the NW PDR of NGC~7023 studied in this paper is characterized by a gas density of $10^{3} - 10^{4}$ $cm^{-3}$ \citep{Chokshi:88, Rogers:95, Fuente:1996, Fuente:1999, Martini:1999, Kohler:2014, Joblin:18} and a radiation field strength, 2600 $G_{0}$\footnote[1]{$G_{0}$ is the intensity of the radiation field in units of the average interstellar radiation field (the Habing field = $1.6 \times 10^{-3}$ erg cm$^{-2}$ s$^{-1}$).} \citep{Chokshi:88, Rogers:95, Joblin:18}, the PDR environments studied in NGC~2023 are characterized by a gas density of $10^{4}$ $cm^{-3}$ \citep{Steiman-Cameron:97, Burton:98, Sandell:15} and a radiation field strength of $10^{3}$ $G_{0}$ \citep{Burton:98, Sandell:15} in the north FOV, and by a gas density of $10^{5}$ $cm^{-3}$ \citep{Steiman-Cameron:97, Sheffer:11, Sandell:15} and a radiation field strength of $10^{4}$ $G_{0}$ \citep{Steiman-Cameron:97, Sheffer:11} in the south FOV. Since the environments studied in NGC~2023 only comprises PDRs, we compared the results of PCA in the PDR of NGC~7023 presented in section 5 to the results of NGC~2023.

The key finding is that in both the nebulae, only two PCs are required to explain the variance in the PAH emission in the PDRs. In both studies, the first PC represented the PAH emission of a mixture of PAHs of neutrals and cations, and the second PC probed the ionization state of PAHs. This indicates that in the PDR environment, the amount of PAH emission and the degree of ionization drive the variation of PAH emission. In Fig.~\ref{fig:characteristic_spec_comparison}, we compare the characteristic PAH spectrum of $PC_{1}$ for NGC~2023 and the PDR environment of NGC~7023. We note that the 6.2, 7.7, 8.6, and 11.2 $\mu$m bands emerge as strong features in both nebulae. However, the relative strength of these features varies between both nebulae. While the strength of the 6.2, 7.7, and 11.2 $\mu$m bands is higher in NGC~7023, the strength of the 8.6 $\mu$m band is similar and the strength of the 11.0 $\mu$m band is higher in NGC~2023 compared to NGC~7023.  This indicates that the PAH population in NGC~2023 may be different from that in NGC~7023. These findings appear to be inconsistent with the grandPAH hypothesis \citep{Andrews:2015}, according to which only a few stable PAHs comprise the astronomical PAH family.

Furthermore, we identified the different character of the subsets of ionic bands in both environments. In NGC~2023, we found that the 6.2 and 7.7 $\mu$m formed one group of ionic bands and the 8.6 and 11.0 $\mu$m the other. While the PCA analysis of PAH emission in NGC~7023 confirmed that the 6.2 and 7.7 $\mu$m bands indeed behave as one group as opposed to the 8.6 and 11.0 $\mu$m bands, we find a further subtle distinction in the behaviour of the 8.6 and 11.0 $\mu$m bands. We have discussed various possibilities for the origin of the different subsets of the ionic bands in Section~\ref{sec:ionic_bands}.

\section{Conclusion}
We have presented the results of a PCA of the fluxes of five PAH bands at 6.2, 7.7, 8.6, 11.0, and 11.2 $\mu$m in NGC~7023. The region of NGC~7023 studied in this paper comprises the NW PDR and the dust-free cavity. We find that only two parameters (PCs) are required to explain most of the observed variance ($\sim$ 98\%) in the PAH fluxes. The first PC ($PC_{1}$), accounting for $\sim$ 84\% of the variance, represents the PAH emission of a mixture of neutral and ionized PAHs and hence the amount of PAH emission. The second PC ($PC_{2}$), accounting for $\sim$ 14\% of the variance, probes the ionization state of PAHs across the nebula. Based on the biplots and correlations of PCs with the PAH ratios, we found that there are subsets of the ionic bands with the 6.2 and 7.7 $\mu$m bands forming one subset and the 8.6 and 11.0 $\mu$m bands the other. The subset comprising the 8.6 and 11.0 $\mu$m bands further shows subtle distinctions in their behaviour.

In addition, PCA analysis shows that the PAH emission characteristics are distinct between the cavity and the PDR. The PAH population in these two regions differs mainly in the ionization state of PAHs, with the PDR consisting of a higher fraction of neutral PAHs than the cavity, which contains more contribution from ionized molecules. The cavity and the PDR also differ in terms of the photochemical evolution of PAHs. While ionization drives PAH variations in the PDR, another molecular property, such as e.g. hydrogenation or molecular edge structure, is responsible for PAH variations in the cavity. Furthermore, we find that the subsets of the ionic bands can only be discerned in the PDR and not in the cavity.

We discussed two likely scenarios for the origin of the subsets of the ionic PAH bands. In one scenario, we argue that the 6.2 and 7.7 $\mu$m bands are less ionized than the 8.6 and the 11.0 $\mu$m bands, thereby resulting in the two distinct subsets of ionic bands. Alternatively, there could be a contribution from VSGs and PAH clusters to the PAH bands in the 6-9 $\mu$m region. In this scenario, the amount of contribution from VSGs and PAH clusters to the PAH bands would lead to the distinct behaviour of the ionic bands in the PDRs.

Finally, we compared the PCA results of PAH emission in NGC~7023 to a similar study conducted previously on NGC~2023. The comparison shows that only two parameters, the amount of PAH emission and the ionization state, drive the variation of PAH emission in PDR-like environments.

\label{sec:Summary}

\section*{Data Availability}
The data underlying this article will be shared on reasonable request to the corresponding author.

\section*{Acknowledgements}
EP and JC acknowledge support from an NSERC Discovery Grant.

\bibliographystyle{apj}
\bibliography{PAH_papers}

\appendix
\appendixpage
\addappheadtotoc

\begin{appendices}
\section{Uncertainty in the coefficients of PCs}
\label{sec:uncertainties}
In Table~\ref{tab:unvertainty_interval}, we provide the upper and the lower limits on the coefficients of the standardized flux variables in PCs estimated by the Monte Carlo Simulation for the PCA performed on the entire FOV.

\begin{table*}
    \centering
    \begin{tabular}{c c c c c c}
    \hline
    \multirow{2}{*}{PC} & \multicolumn{5}{c}{Upper and lower limits on coefficients} \\
    & $z_{6.2}$ & $z_{7.7}$ & $z_{8.6}$ & $z_{11.0}$ & $z_{11.2}$ \\
    \hline
    1 & 0.481 -- 0.484 & 0.475 -- 0.480 & 0.473 -- 0.477 & 0.452 -- 0.461 & 0.321 -- 0.326 \\ 
    2 & 0.107 -- 0.128 & -0.163 -- -0.132 & -0.277 -- -0.246 & -0.358 -- -0.296 & 0.880 -- 0.893\\
    3 & -0.396 -- -0.339  & -0.575 -- -0.508 & 0.012 -- 0.185 & 0.648 -- 0.751 & 0.207 -- 0.274\\
    4 & -0.699 -- -0.071 & -0.413 -- 0.647 & -0.198 -- 0.842 & -0.422 -- 0.133& 0.041 -- 0.188\\
    5 & -0.784 -- -0.331 & -0.039 -- 0.541 & 0.000 -- 0.794& -0.499 -- 0.019 & 0.097 -- 0.217\\
    \hline
    \end{tabular}
    \caption{The upper and lower limits on the coefficients of the standardized flux variables in PCs estimated by the Monte Carlo Simulation for the PCA performed on the entire FOV.}
    \label{tab:unvertainty_interval}
\end{table*}

\section{Spatial Map of Total PAH flux}
\label{sec:Spatial_map_PAH_flux}
\begin{figure}
\centering
\includegraphics[angle=265.11918,scale=0.45]{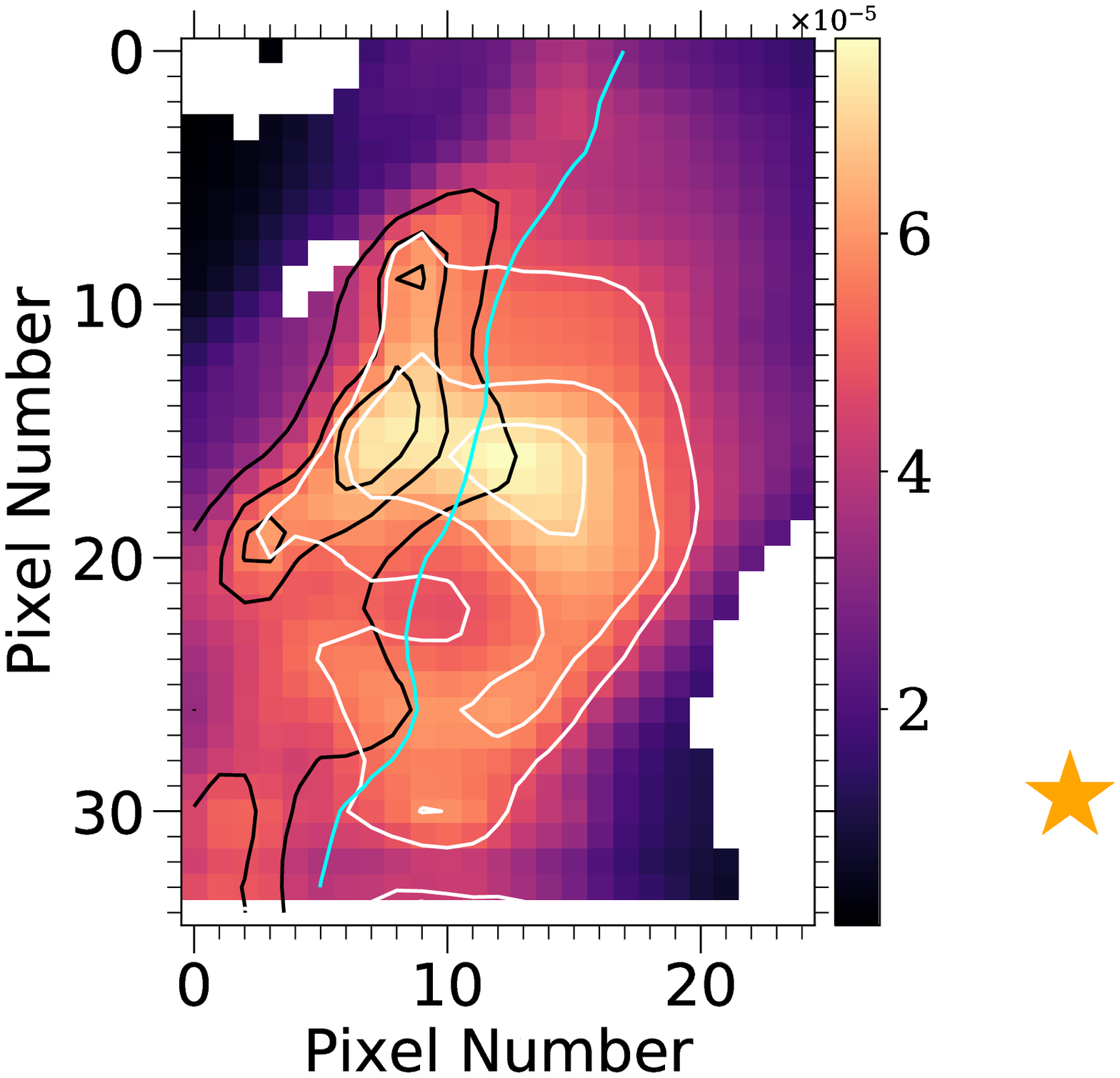}
 
\caption{Spatial map of total PAH flux (the sum of the fluxes of PAH bands at 6.2, 7.7, 8.6, 11.0, and 11.2 $\mu$m). For reference, the 7.7, 11.2 $\mu$m PAH intensity, and $PC_{2}$ = -0.24, are represented by white, black, and cyan contours respectively. The yellow star outside the FOV indicates the position of the illuminating star.}
\label{fig:spatial_map_total_PAH_flux}

\end{figure}
In Fig.~\ref{fig:spatial_map_total_PAH_flux}, we show the spatial distribution of the total PAH flux, i.e. the sum of the fluxes of PAH bands at 6.2, 7.7, 8.6, 11.0, and 11.2 $\mu$m in the SL FOV  of NGC~7023. 

\section{Distribution of the PAH ratios in the PDR}
\label{sec:distribution_PAH_ratios_PDR}
\begin{figure*}
\centering

\includegraphics[scale=0.3]{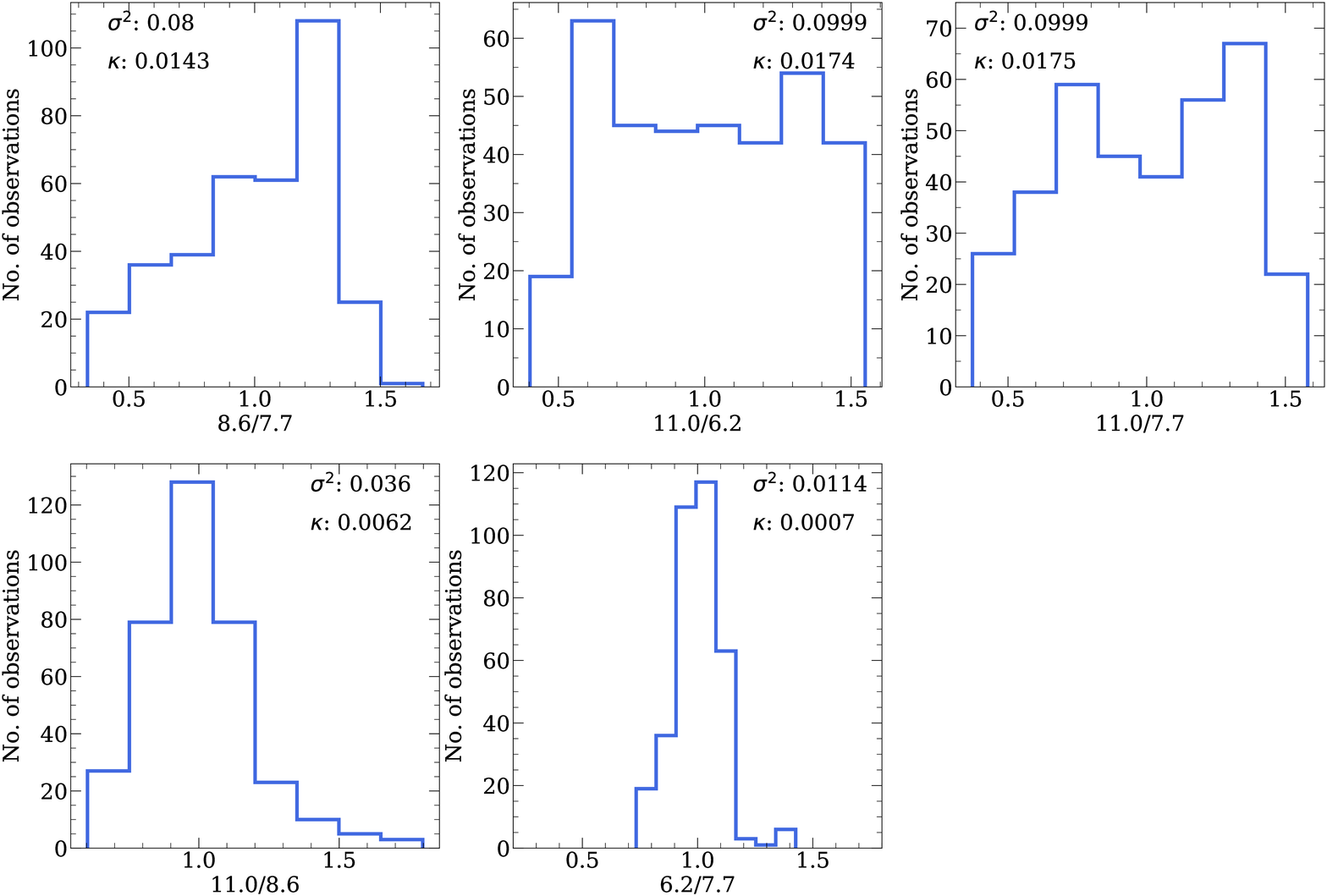}
 
\caption{Distribution of PAH ratios normalized to their mean value in the PDR of NGC 7023. The second moment, $\sigma^{2}$, and the fourth moment, $\kappa$, are shown in the corner of each plot.}
\label{fig:distribution_PAH_ratios}

\end{figure*}

In Fig.~\ref{fig:distribution_PAH_ratios}, we show the distribution of the PAH ratios in the PDR. To facilitate the comparison between the distributions of various PAH ratios, we show the distribution of the normalized PAH ratios obtained by dividing the ratios by their mean value. We also calculated the second moment, i.e. the variance ($\sigma^{2}$), and the fourth moment, i.e. the kurtosis ($\kappa$) of the distribution. The $\sigma^{2}$ is a measure of the width of the distribution from the mean value. The $\kappa$ is a measure of the heaviness of the tails of the distribution. We note that the 6.2/7.7 values exhibit the lowest $\sigma^{2}$ and $\kappa$ values followed by the 11.0/8.6 values.

\section{PCA of PAH emission in the PDR}
\label{sec:PCA_PAH_PDR}

\begin{figure*}
\centering
\begin{tabular}{cc}
    \includegraphics[angle=265.11918, scale=0.45]{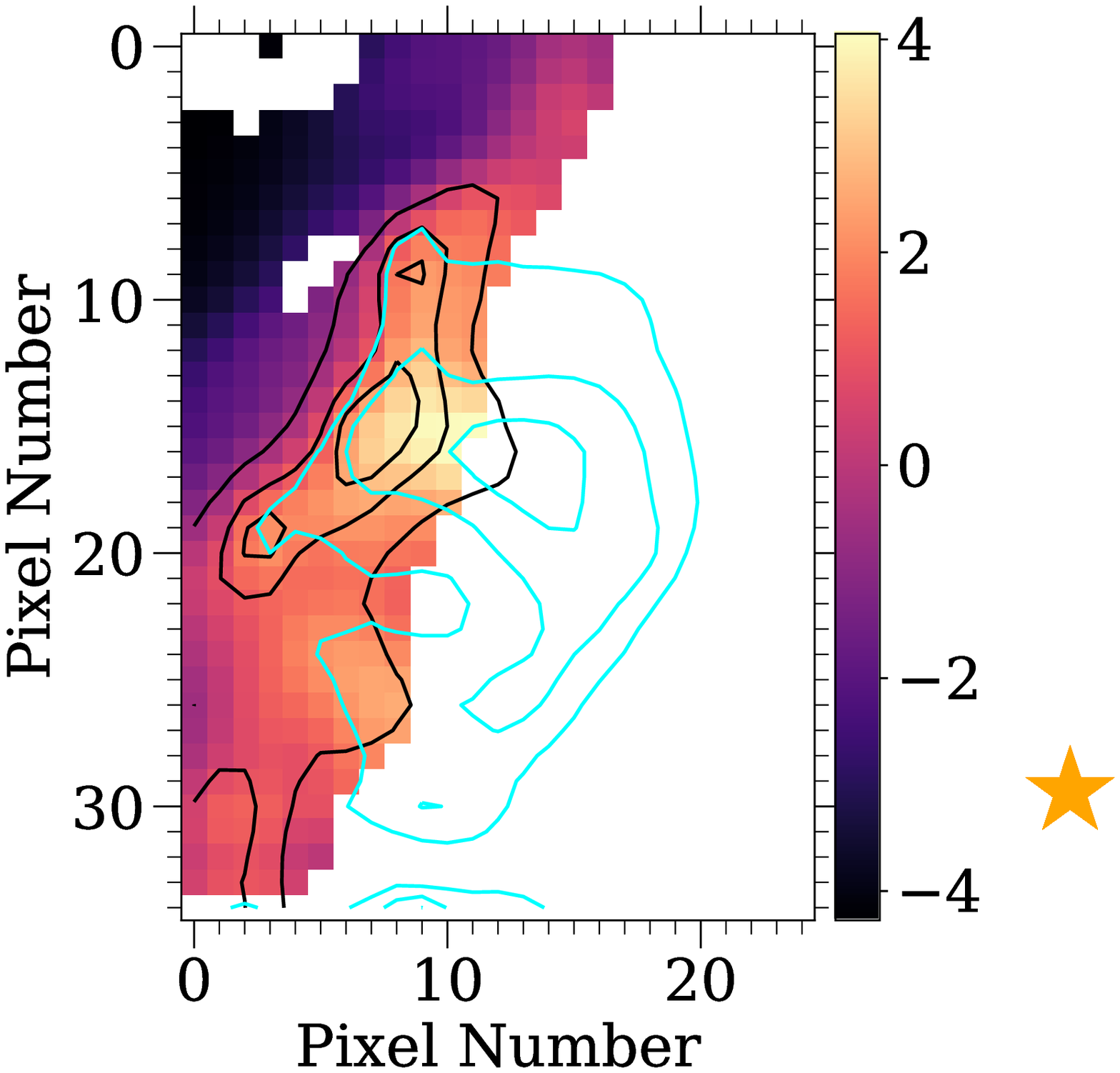} & \includegraphics[angle=265.11918, scale=0.45]{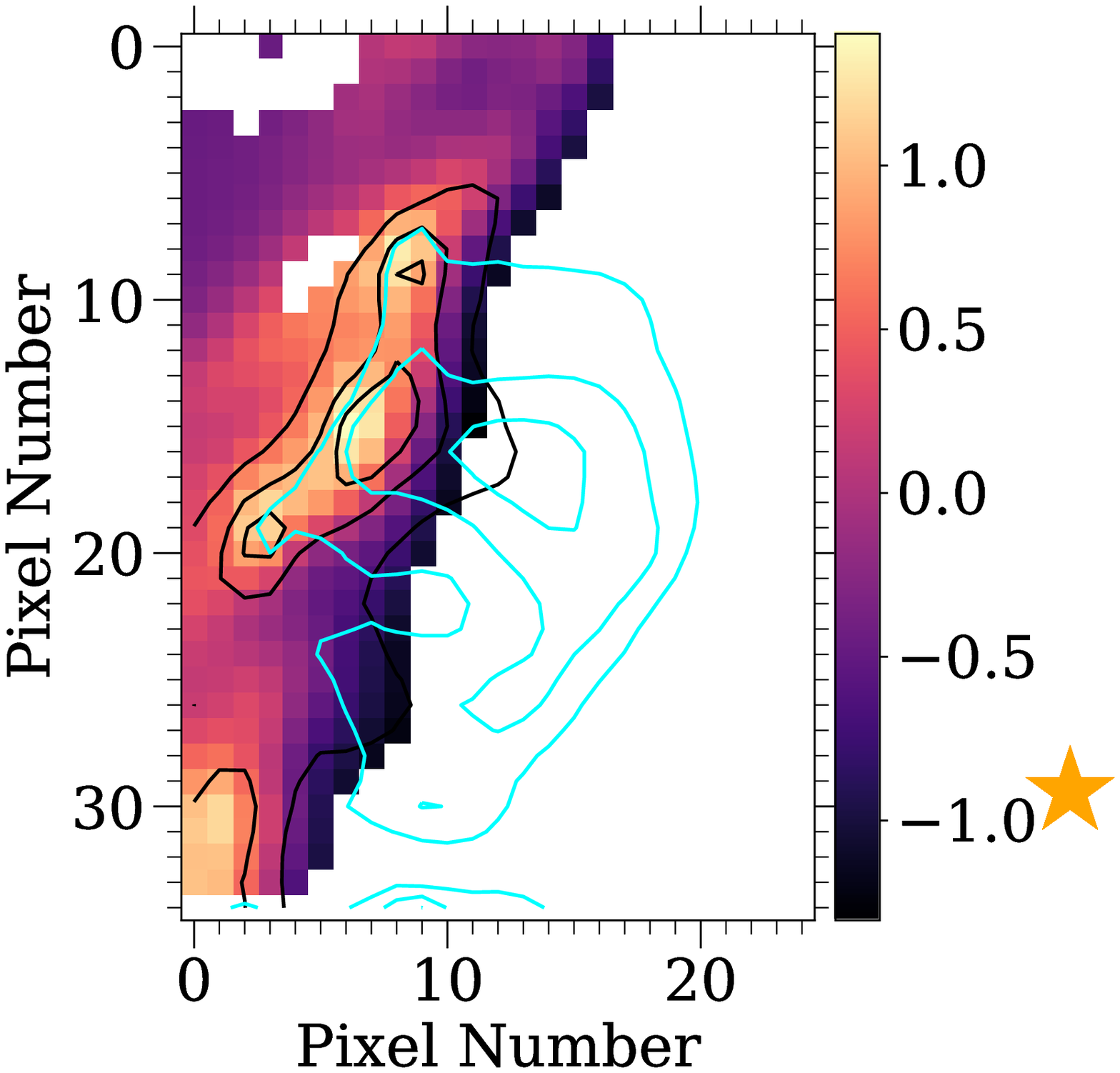}  \\
     & 
\end{tabular}
 
\caption{Spatial maps of $PC_{1}$ (left) and $PC_{2}$ (right) obtained from the independent PCA analysis of PAH emission in the PDR of NGC~7023. For reference, the contours of the 7.7 and 11.2 $\mu$m PAH intensity are overlaid in cyan and black respectively. The yellow star outside the FOV indicates the position of the illuminating star.}
\label{fig:spatial_map_PCs_PDR}

\end{figure*}

\begin{table}
    \centering
    
    \begin{tabular}{c c c}
    \hline
    \multirow{2}{*}{PAH band} & $\langle I_{PAH}\rangle$ & $\sigma_{PAH}$  \\
    & ($\times 10^{-5}$) & ($\times 10^{-6}$)\\
    \hline
    6.2 & 1.212 & 5.390  \\ 
    7.7 & 1.846 & 8.199 \\
    8.6 & 0.321  & 1.855 \\
    11.0 & 0.008  &  0.049\\
    11.2 & 0.643  & 3.010 \\
    \hline
    \end{tabular}
    \caption{The mean ($\langle I_{PAH}\rangle$) and standard deviation ($\sigma_{PAH}$) values of the PAH band flux variables in the PDR of NGC~7023. All values are in units of ${\rm W m}^{-2}{\rm sr}^{-1}$.}
    \label{tab:stats_flux_PDR}
\end{table}

We  performed an independent PCA of PAH band flux variables observed in the NW PDR of NGC~7023 (see \ref{sec:cavity_PDR} for details). Table~\ref{tab:stats_flux_PDR} lists the mean and standard deviation values of the PAH band flux variables. The PC eigenvectors that result from this PCA are
\begin{equation}
\begin{split} 
PC_{1} = & \,\,  0.462\, z_{6.2} + 0.461\, z_{7.7} + 0.457\, z_{8.6}  \\
& + 0.434\, z_{11.0} + 0.420\, z_{11.2}\\
PC_{2} = & \,\,0.237\, z_{6.2} + 0.028\, z_{7.7} - 0.345\,z_{8.6}  \\
& - 0.588\, z_{11.0} + 0.691\, z_{11.2}\\
PC_{3} =&\,\, - 0.257\, z_{6.2} - 0.687\, z_{7.7} + 0.067\,z_{8.6}  \\
& + 0.434\,z_{11.0} + 0.519\,z_{11.2}\\
PC_{4} =&\,\,- 0.411\, z_{6.2} + 0.489\, z_{7.7} - 0.625\,z_{8.6}  \\
& + 0.420\,z_{11.0} + 0.167\, z_{11.2}\\
PC_{5} =&\,\, - 0.703\, z_{6.2} + 0.281\, z_{7.7} + 0.526\,z_{8.6}  \\
& - 0.318\,z_{11.0} + 0.221\,z_{11.2}
\end{split}
\label{eq:PC_PDR}
\end{equation}

In Table~\ref{tab:unvertainty_interval_PDR}, we provide the upper and the lower limits on the coefficients of the standardized flux variables in PCs estimated by the Monte Carlo Simulation.

\begin{table*}
    \centering
    \begin{tabular}{c c c c c c}
    \hline
    \multirow{2}{*}{PC} & \multicolumn{5}{c}{Upper and lower limits on coefficients} \\
    & $z_{6.2}$ & $z_{7.7}$ & $z_{8.6}$ & $z_{11.0}$ & $z_{11.2}$ \\
    \hline
    1 & \,\,0.462 -- \,\,0.467 & \,\,0.461 -- \,\,0.467 & \,\,0.455 -- \,\,0.462 & \,\,0.417 -- \,\,0.434 & 0.420 -- 0.424 \\ 
    2 & \,\,0.230 -- \,\,0.249 & \,\,0.012 -- \,\,0.068 & -0.345 -- -0.279 & -0.675 -- -0.588 & 0.634 -- 0.694\\
    3 & -0.257 -- -0.182  & -0.687 -- -0.555 & -0.254 -- \,\,0.067 & \,\,0.434 -- \,\,0.574 & 0.498 -- 0.578\\
    4 & -0.411 -- \,\,0.003 & -0.498 -- \,\,0.489 & -0.625 -- \,\,0.826 & -0.440 -- \,\,0.420 & 0.012 -- 0.167\\
    5 & -0.824 -- -0.703 & \,\,0.281 -- \,\,0.512 & \,\,0.029 -- \,\,0.526 & -0.318 -- \,\,0.022 & 0.221 -- 0.292\\
    \hline
    \end{tabular}
    \caption{The upper and lower limits on the coefficients of the standardized flux variables in PCs estimated by the Monte Carlo Simulation for the independent PCA analysis of PAH emission in the PDR of NGC~7023. }
    \label{tab:unvertainty_interval_PDR}
\end{table*}

We can further decompose the standardized flux variables in the NW PDR of NGC~7023 into two most important PCs ($PC_{1}$ and $PC_{2}$) as follows
\begin{equation}
\begin{split} 
z_{6.2} = & \,\,  0.462\, PC_{1} + 0.237\, PC_{2} \\
z_{7.7} = & \,\,0.461\, PC_{1} + 0.028\, PC_{2} \\
z_{8.6} =&\,\, 0.457\, PC_{1} - 0.345\, PC_{2} \\
z_{11.0} =&\,\, 0.434\, PC_{1} - 0.588\, PC_{2} \\
z_{11.2} =&\,\, 0.420\, PC_{1} + 0.691\, PC_{2} 
\end{split}
\label{eq:std_var_decomposition_PDR}
\end{equation}

In Fig.~\ref{fig:spatial_map_PCs_PDR}, we present the spatial maps of PCs obtained for this PCA of PAH emission in the PDR.

\section{PCA of PAH emission in the cavity}
\label{sec:PCA_PAH_cavity}

\begin{figure*}
\centering
\begin{tabular}{cc}
    \includegraphics[angle=265.11918, scale=0.45]{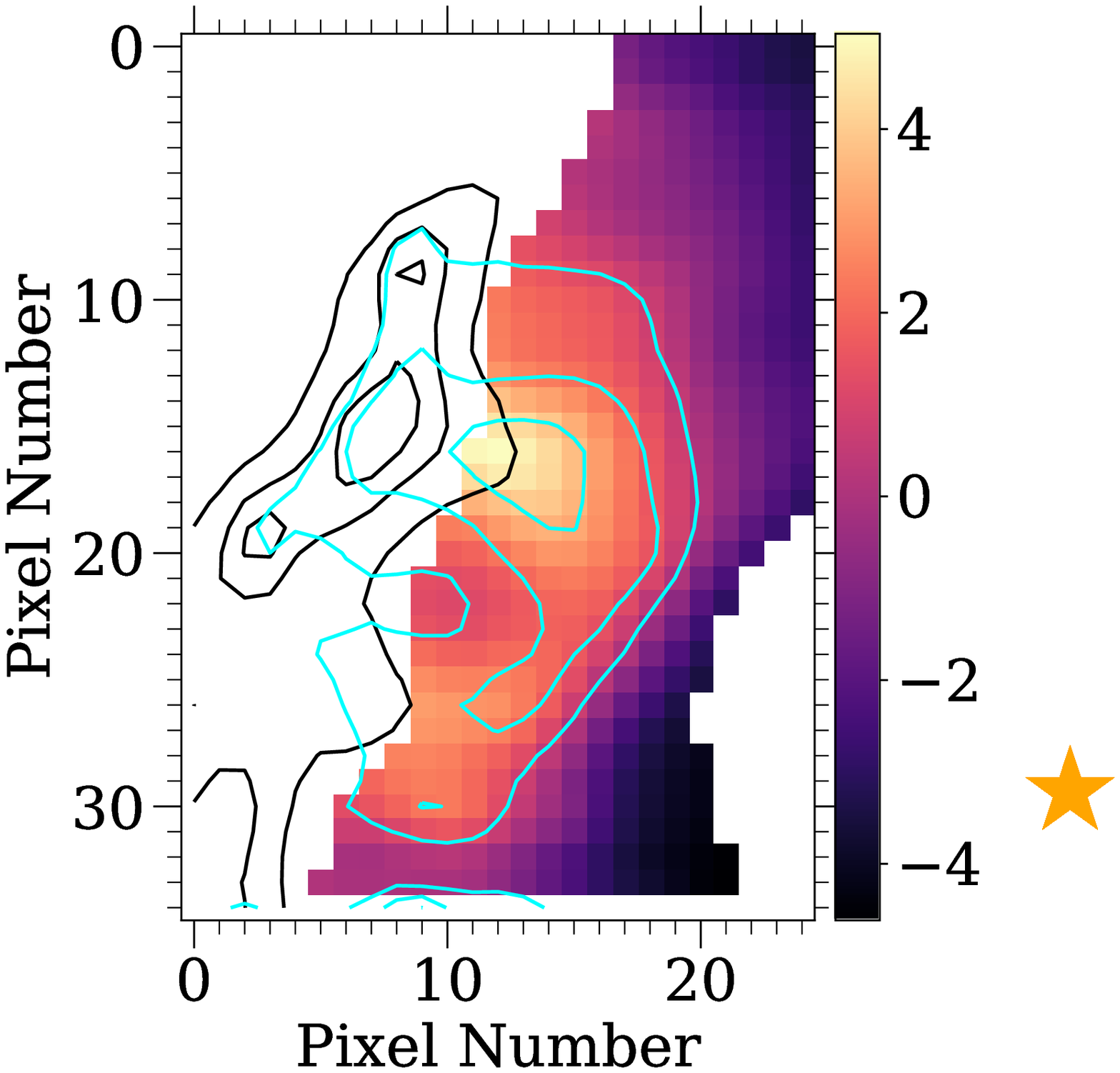} & \includegraphics[angle=265.11918, scale=0.45]{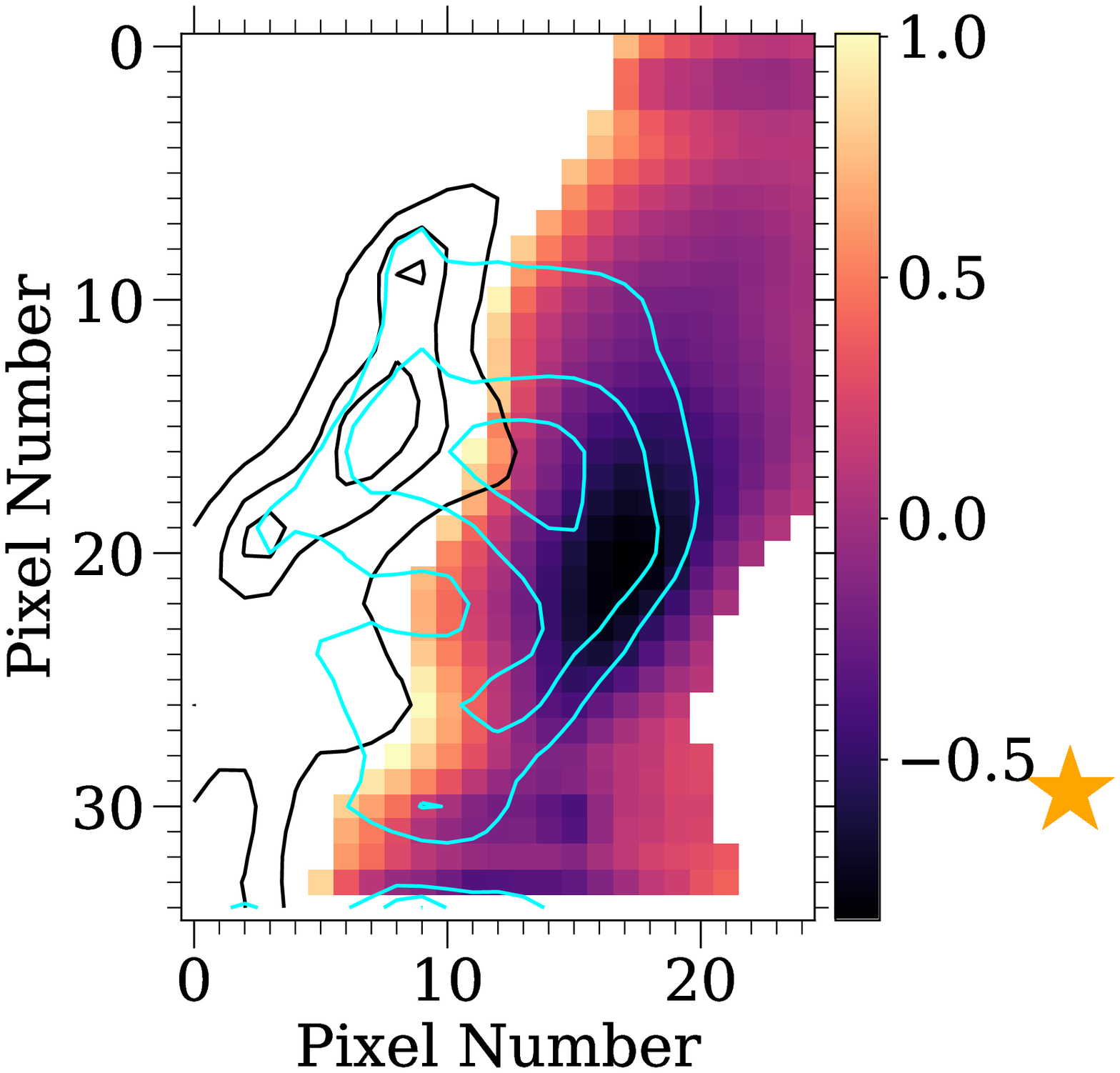}  \\
     & 
\end{tabular}
 
\caption{Spatial maps of $PC_{1}$ (left) and $PC_{2}$ (right) obtained from the independent PCA analysis of PAH emission in the cavity of NGC~7023. For reference, the contours of the 7.7 and 11.2 $\mu$m PAH intensity are overlaid in cyan and black respectively. The yellow star outside the FOV indicates the position of the illuminating star.}
\label{fig:spatial_map_PCs_cavity}

\end{figure*}

\begin{table}
    \centering
    
    \begin{tabular}{c c c}
    \hline
    \multirow{2}{*}{PAH band} & $\langle I_{PAH}\rangle$ & $\sigma_{PAH}$  \\
    & ($\times 10^{-5}$) & ($\times 10^{-6}$)\\
    \hline
    6.2 & 1.237 & 4.751  \\ 
    7.7 & 2.200 & 7.971 \\
    8.6 & 0.420  & 1.695 \\
    11.0 & 0.011  &  0.044 \\
    11.2 & 0.305  & 1.612 \\
    \hline
    \end{tabular}
    \caption{The mean ($\langle I_{PAH}\rangle$) and standard deviation ($\sigma_{PAH}$) values of the PAH band flux variables in the cavity of NGC~7023. All values are in units of ${\rm W m}^{-2}{\rm sr}^{-1}$.}
    \label{tab:stats_flux_cavity}
\end{table}

We also performed a PCA of PAH band flux variables in the cavity of NGC~7023. Table~\ref{tab:stats_flux_cavity} lists the mean and standard deviation values of the variables used in the PCA. The PC eigenvectors that result from this PCA are
\begin{equation}
\begin{split} 
PC_{1} = & \,\,  0.451\, z_{6.2} + 0.446\, z_{7.7} + 0.454\, z_{8.6}  \\
& + 0.450\, z_{11.0} + 0.433\, z_{11.2}\\
PC_{2} = & \,\,- 0.326\, z_{6.2} - 0.481\, z_{7.7} - 0.112\,z_{8.6}  \\
& + 0.154\, z_{11.0} + 0.791\, z_{11.2}\\
PC_{3} =&\,\,  -0.029\, z_{6.2} + 0.579\, z_{7.7} - 0.314\,z_{8.6}  \\
& - 0.626\,z_{11.0} + 0.417\,z_{11.2}\\
PC_{4} =&\,\,  0.831\, z_{6.2} - 0.413\, z_{7.7} - 0.294\,z_{8.6}  \\
& - 0.212\,z_{11.0} + 0.091\, z_{11.2}\\
PC_{5} =&\,\, - 0.007\, z_{6.2} - 0.254\, z_{7.7} + 0.772\,z_{8.6}  \\
& - 0.578\,z_{11.0} + 0.065\,z_{11.2}
\end{split}
\label{eq:PC_cavity}
\end{equation}

In Table~\ref{tab:unvertainty_interval_cavity}, we provide the upper and the lower limits on the coefficients of the standardized flux variables in PCs estimated by the Monte Carlo Simulation.

\begin{table*}
    \centering
    \begin{tabular}{c c c c c c}
    \hline
    \multirow{2}{*}{PC} & \multicolumn{5}{c}{Upper and lower limits on coefficients} \\
    & $z_{6.2}$ & $z_{7.7}$ & $z_{8.6}$ & $z_{11.0}$ & $z_{11.2}$ \\
    \hline
    1 & \,\,0.451 -- \,\,0.453 & \,\,0.446 -- \,\,0.448 & \,\,0.454 -- \,\,0.456 & \,\,0.446 -- \,\,0.452 & 0.432 -- 0.436 \\ 
    2 & -0.353 -- -0.309 & -0.507 -- -0.456 & -0.162 -- -0.090 & \,\,0.074 -- \,\,0.306 & 0.714 -- 0.821\\
    3 & -0.029 -- \,\,0.194  & \,\,0.157 -- \,\,0.579 & -0.314 -- \,\,0.129 & -0.878 -- -0.626 & 0.350 -- 0.529\\
    4 & -0.465 -- \,\,0.831 & -0.413 -- \,\,0.699 & -0.783 -- -0.294 & -0.212 -- \,\,0.240 & 0.091 -- 0.179\\
    5 & -0.007 -- \,\,0.819 & -0.410 -- -0.124 & -0.720 -- \,\,0.772 & -0.578 -- \,\,0.095 & 0.026 -- 0.101\\
    \hline
    \end{tabular}
    \caption{The upper and lower limits on the coefficients of the standardized flux variables in PCs estimated by the Monte Carlo Simulation for the independent PCA analysis of PAH emission in the cavity of NGC~7023.}
    \label{tab:unvertainty_interval_cavity}
\end{table*}

We can decompose the standardized flux variables in the cavity of NGC~7023 into two most important PCs ($PC_{1}$ and $PC_{2}$) as follows
\begin{equation}
\begin{split} 
z_{6.2} = & \,\,  0.451\, PC_{1} - 0.326\, PC_{2} \\
z_{7.7} = & \,\,0.446\, PC_{1} - 0.481\, PC_{2} \\
z_{8.6} =&\,\, 0.454\, PC_{1} - 0.112\, PC_{2} \\
z_{11.0} =&\,\, 0.450\, PC_{1} + 0.154\, PC_{2} \\
z_{11.2} =&\,\, 0.433\, PC_{1} + 0.791\, PC_{2} 
\end{split}
\label{eq:std_var_decomposition_cavity}
\end{equation}

In Fig.~\ref{fig:spatial_map_PCs_cavity}, we present the spatial maps of PCs obtained for this PCA of PAH emission in the cavity.

\end{appendices}
\bsp	
\label{lastpage}
\end{document}